\documentclass[%
 reprint,
 amsmath,amssymb,
 aps,
floatfix,
]{revtex4-1}
\usepackage{amsmath}
\usepackage{notes2bib}
\usepackage{soul}
\usepackage{graphicx}
\usepackage{dcolumn}
\usepackage{bm}
\usepackage[T1]{fontenc}
\usepackage[dvipsnames]{xcolor}
\newcommand{\authormain}{Farshid Jafarpour}
\newcommand{\titlemain}{Asymptotic decoupling of growth rate and cell-size distribution}
\usepackage[bookmarks={false}, pdfauthor={\authormain}, pdftitle={\titlemain}]{hyperref}
\hypersetup{colorlinks=true, linkcolor=BrickRed, citecolor=Violet, filecolor=OliveGreen, urlcolor=RoyalBlue, filebordercolor={.8 .8 1}, urlbordercolor={.8 .8 0}}

\newcommand{\eq}[1]{Eq.~\eqref{eq:#1}}
\newcommand{\Eq}[1]{Equation~\eqref{eq:#1}}
\newcommand{\fig}[1]{Fig.~\ref{fig:#1}}
\newcommand{\Fig}[1]{Figure~\ref{fig:#1}}
\newcommand{\rf}[1]{Ref.~\cite{#1}}
\newcommand{\Rf}[1]{Reference~\cite{#1}}

\begin{document}

\title{Asymptotic decoupling of population growth rate and cell size distribution}

\author{Ya\"ir Hein}
\affiliation{
	Institute for Theoretical Physics,
	Utrecht University, 3584 CC Utrecht, Netherlands}
\author{Farshid Jafarpour}%
\affiliation{
	Institute for Theoretical Physics,
	Utrecht University, 3584 CC Utrecht, Netherlands}

\date{\today}

\begin{abstract}
    The rate at which individual bacterial cells grow depends on the concentrations of cellular components such as ribosomes and proteins. These concentrations continuously fluctuate over time and are inherited from mother to daughter cells, leading to correlations between the growth rates of cells across generations. Division sizes of cells are also stochastic and correlated across generations due to a phenomenon known as cell size regulation. Fluctuations and correlations from both growth and division processes affect the population dynamics of an exponentially growing culture. Here, we provide analytic solutions for the population dynamics of cells with continuously fluctuating growth rates coupled with a generic model of cell-size regulation. We show that in balanced growth, the effects of growth and division processes decouple; the population growth rate only depends on the single-cell growth rate process, and the population cell size distribution only depends on the model of division and cell size regulation. The population growth rate is always higher than the average single-cell growth rate, and the difference increases with growth rate variability and its correlation time. This difference also sets the timescale for the population to reach its steady state. We provide analytical solutions for oscillations in population growth rate and traveling waves in size distribution during this approach to the steady state.
\end{abstract}
\maketitle

\section{Introduction}
For over a century, bacteria have been studied in population experiments in which the number of cells grows exponentially. 
In the past decade, however, numerous experiments have attempted to uncover models of bacterial growth, division, and cell size control at the single-cell level~\cite{stewart2005aging,wang2010robust,campos2014constant,robert2014division,taheri2015cell,tanouchi2017long,kar2021distinguishing,vashistha2021non,golubev2016applications,nordholt2020biphasic}. The details of these single-cell models ultimately determine the dynamics of an exponentially growing population and its population-level observables, such as cell size distribution and population growth rate, the latter of which is an important indicator of population fitness~\cite{powell1956growth,powell1964note,amir2014cell,hashimoto2016noise,lin2017effects,thomas2018analysis,thomas2018sources,jafarpour2018bridging,jafarpour2019cell,garcia2019linking,vittadello2019mathematical,tuanase2008regulatory,thomas2017single,levien2020interplay,lin2020single,van2017taking,ho2018modeling,nozoe2020cell,genthon2022analytical,barber2021modeling,jia2021cell}.

\begin{figure}[h!]
    \centering
    \includegraphics[width=\columnwidth]{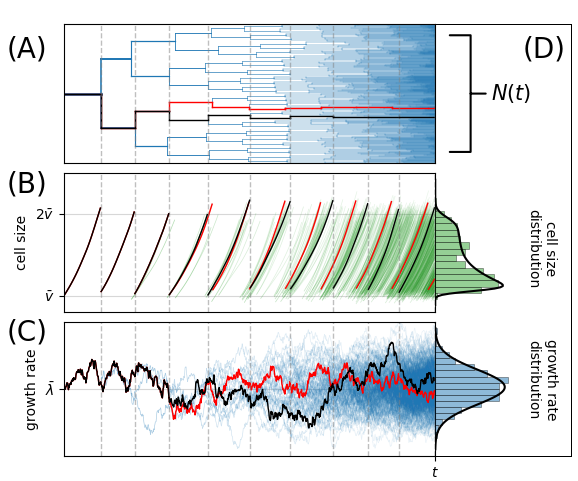}
    \caption{Relationship between population properties and single-cell statistics: (A) Lineage tree of a population, two lineages are highlighted. (B) Cell sizes along each lineage grow approximately exponentially and divide by two at the division. (C) Along each lineage, the growth rate fluctuates over time. (D) The state of the population at time $t$ is characterized by the number of cells and the distributions of their growth rates and sizes. Both the number of cells and their sizes at time $t$ depend on the history of the stochastically branching tree of stochastic differential equations of growth rates in (C). The growth processes of every two lineages in the tree share a random portion of their history before they branch. The positions of these branches (division times) are random and depend on the size of the cell and also the size at the time of the previous branch (birth size) both of which again depend on the history of the growth rate and the number of earlier divisions. To relate the properties of the population to the single-cell statistics using this highly interconnected and growing set of both continuous and discrete stochastic processes is a daunting task that has consistently forced modelers to make unphysical simplifying assumptions such as ignoring fluctuations or correlations in the growth or in the division process~\cite{powell1956growth,powell1964note,amir2014cell,hashimoto2016noise,lin2017effects,thomas2018analysis,thomas2018sources,jafarpour2018bridging,jafarpour2019cell,garcia2019linking,vittadello2019mathematical,tuanase2008regulatory,thomas2017single,levien2020interplay}.}
    \label{fig:tree}
\end{figure}

In constant environments, single bacterial cells grow approximately exponentially in size and divide when they roughly double their sizes. However, the division sizes and single-cell growth rates fluctuate and vary from cell to cell~\cite{wang2010robust,campos2014constant,taheri2015cell,pirjol2017phenomenology}. Moreover, the time it takes for a cell to divide is correlated between a mother cell and its daughter cells. There are two distinct sources of correlations. On one hand, if a cell grows for too long before dividing, its daughter cells would be larger than average and would have to divide earlier to compensate for their size; this process is known as cell size regulation, and it induces negative correlations in inter-division times of the mother and daughter cells. On the other hand, cells inherit the concentrations of proteins and ribosomes (which set their instantaneous growth rate) from their mother cells at the moment of division~\cite{kiviet2014stochasticity,lin2021disentangling,kleijn2018noise,thomas2018sources}. This leads to fast-growing cells having fast-growing daughter cells, which induces positive correlations in inter-division times. These two competing correlations both decay with a comparable timescale of about a couple of generation times~\cite{wang2010robust,taheri2015cell}. Despite numerous theoretical and experimental attempts, the effects of the fluctuations in growth and division and the two competing sources of correlations in inter-division times on population-level observables 
are not well understood (the mathematical difficulty of the relationship between single-cell and population variables is illustrated in \fig{tree}).

In this paper, we introduce a model in which the fluctuations in single-cell growth rates are described by a continuous stochastic differential equation and accurately incorporates noise and correlations from both growth and division. We demonstrate a surprising decoupling of population growth rate and the distribution of cell sizes in the population at the steady state: population cell size distribution is determined by the model of division and contains no information about the single-cell growth process, whereas population growth rate is determined solely by the model of single-cell growth dynamics, and is completely unaffected by details of cell division and cell-size regulation. We show that the population growth rate is always larger than the average single-cell growth rate. This difference is simply the product of two single-cell variables, namely growth rate variability and its correlation time. It follows that both variability and correlations in growth are beneficial to the population fitness when fixing the time-averaged growth rate.

In addition to the steady-state results, we have exact solutions for the population dynamics. Previous works have demonstrated that population properties can oscillate for a long time before reaching a steady state~\cite{chiorino2001desynchronization,vittadello2019mathematical,GAVAGNIN20211314,jafarpour2019cell,brown1940note,kendall1948role,hirsch1966decay,hein2023competition}. Our analytical solutions provide an exact form for the oscillations in the growth rate and the accompanying traveling waves in the cell size distribution. We show that these oscillations decay at a rate proportional to the difference between the population growth rate and the average single-cell growth rate. These oscillations can therefore last for an unexpectedly long time when the correlation time and growth rate variability are small.\\

\section{Model} We assume cell sizes grow exponentially with some growth rate $\lambda_t$, 
\begin{equation}
\label{eq:v_t0}
    \frac{d}{dt} v(t) = \lambda_t v(t),
\end{equation}
at all $t$ where no division is taking place. The growth rate, $\lambda_t$, can be any stationary stochastic process that remains continuous through cell division. This process models how cell growth varies with the internal protein and ribosome concentrations which fluctuate over time and are inherited by the daughter cells upon division~\cite{kiviet2014stochasticity,kleijn2018noise}. 

We assume that an $n$th generation cell divides when its size reaches a random division size $v_{d,n}$ that may depend on its birth size and possibly those of its ancestors through a stochastic function $h$,
\begin{equation}\label{eq:h}
    v_{d,n}= h(v_{b,n},v_{b,n-1},\dots).
\end{equation}
The function $h$ defines the model of division and cell size regulation and determines the steady-state birth size distribution $f_{v_b}$. This generalized formulation is compatible with many known models of cell-size control (see Refs.~\cite{amir2014cell,ho2018modeling} for examples of such models).

A specific example of such a model of growth and division is as follows.
Growth rate $\lambda_t$ can be an Ornstein-Uhlenbeck process, characterized by the differential equation
\begin{equation}
\label{eq:l_ou}
    \frac{d\lambda_t}{dt}= -\theta(\lambda_t-\bar{\lambda}) + \sigma_{\lambda}\sqrt{2\theta}\,\xi(t),
\end{equation}
where $\bar\lambda$ is the mean growth rate, $\sigma^2_\lambda$ is the growth rate variance, $\xi(t)$ is standard Gaussian white noise and $\theta$ sets the auto-correlation decay rate ($1/\theta$ is the correlation time). It follows that the growth rate is Gaussian distributed with the covariance $\text{Cov}(\lambda_t,\lambda_s)=\sigma_\lambda^2  e^{-\theta|t-s|}$. Many naturally occurring random processes that fluctuate around some equilibrium value are well-approximated by this process. An example of this process is given in \fig{tree}(C).

An example of a division model of the type \eq{h} is based on \rf{amir2014cell}, where cell division size $v_d$ is determined based on its birth size $v_b$
\begin{equation}
    \label{eq:csr}
    \ln(v_d) = \alpha\ln(2\bar v) + (1-\alpha)\ln(2v_b) + \eta,
\end{equation}
where $\eta$ is a Gaussian random variable with mean zero and variance $\sigma^2_\eta$, $\bar{v}$ is the constant equilibrium birth size, and $\alpha$ determines the degree of cell size regulation ($\alpha=0$ is the timer model, $\alpha=1$ is the sizer model, and $\alpha=0.5$ approximates the adder model; see \rf{amir2014cell,ho2018modeling} for more details). 
\\

We will first derive the population growth rate and the size distribution for a steady-state population based on \eq{l_ou} and \eq{csr}. Interestingly, it turns out the population cell size distribution depends only on the division parameters and the population growth rate depends only on the single cell growth rate parameters. We then show that this decoupling holds more generally by deriving the steady-state population growth rate and cell-size distribution for an arbitrary growth rate $\lambda_t$ and division models of the form \eq{h}. Finally, we discuss the time-dependent behavior of a population out of steady-state and give the full time-dependent population growth and cell size distribution for the specific model based on \eq{l_ou} and \eq{csr}.

\section{Derivation of Population Properties}
\label{sec:Derivation}
In this section, we assume that the growth rate $\lambda_t$ is given by an Ornstein-Uhlenbeck process of the form in \eq{l_ou}.
We start our derivation of the population properties by defining a dimensionless stochastic time as
\begin{equation}
\label{eq:tau_def0}
    \tau_t \equiv \int_0^t \lambda_s ds.
\end{equation}
Since $\lambda_t$ is Gaussian, $\tau_t$ is also Gaussian. For large $t$, the mean and variance become linear in $t$ given by
\begin{equation}
\label{eq:tau_mean}
    \langle \tau_t\rangle \approx \bar{\lambda} t + \frac{1}{\theta}(\lambda_0-\bar\lambda), \text{ and}\quad \sigma_{\tau_t}^2 \approx
    2\frac{\sigma_\lambda^2}{\theta}t- 3\frac{\sigma_\lambda^2}{\theta^2}.
\end{equation}
Explicit expressions for $\langle \tau_t\rangle$ and $\sigma_{\tau_t}^2$ with their derivations are given in Appendix \ref{sec:OU process}. The dynamics of $v(\tau)$ in terms of stochastic time are simplified to
\begin{equation}
\label{eq:v_tau0}
    \frac{d}{d\tau} v(\tau)= v(\tau).
\end{equation}
This follows from the chain rule of differentiation and the fact that $d\tau_t/dt=\lambda_t$. By doing this, we absorb all growth fluctuations into the stochastic time, resulting in cell size dynamics where the only stochasticity comes from the division noise. \Eq{v_tau0} now implies that
\begin{equation} 
\label{eq:vrel}
    v(\tau)= \frac{v_0\,e^{\tau}}{2^{\Delta (\tau)}},
\end{equation}
where $\Delta(\tau)$ is the number of divisions before $\tau$.
Here $v_0e^\tau$ corresponds to the size the cell would have at $\tau$ if no divisions had taken place, and the factor of $1/2^{\Delta(\tau)}$
corresponds to the cell having halved its size for a total of $\Delta(\tau)$ times.

\Rf{jafarpour2019cell} shows that in the absence of growth fluctuations, all the population properties are periodic in time. Since all the growth fluctuations are absorbed into $\tau$, the distribution of $v(\tau)$ would also be periodic with a period of $\ln(2)$ (see Appendix~\ref{sec:Periodicity} for a mathematical proof).

We can now relate the lineage behavior to the population behavior. Our first focus is to find the expected population size $N(t)$ at some time $t$. There are $2^n$ distinct $n$-th generation cells in the population tree, each of which can be associated with a unique lineage. These lineages each have the same probability $P(\Delta(\tau_t)=n)$ of being in the $n$-th generation at time $t$. Hence, the expected number of $n$-th generation cells present at time $t$ is given by $2^nP(\Delta(\tau_t)=n)$. The total expected number of cells is the sum of this quantity over all generations~\cite{levien2020large,pigolotti2021generalized,jafarpour2019cell}
\begin{equation}
\label{eq:Ndel}
    N(t) = \sum_{n=0}^\infty 2^n P(\Delta(\tau_t)=n) = \left\langle 2^{\Delta (\tau_t)} \right\rangle.
\end{equation}
Since divisions are triggered purely by cell size, $\Delta(\tau_t)$ depends only on $\tau_t$ and not directly on $t$. We may, therefore, average over $\Delta(\tau_t)$ and $\tau_t$ independently,
\begin{equation}
\label{eq:N_hat0}
    N(t) = \left\langle \left\langle\left. 2^{\Delta(\tau_t)}\right| \tau_t\right\rangle\right\rangle.
\end{equation}

By applying \eq{vrel}, we can rewrite the inner expected value conditioned to $\tau_t=\tau$ as 
\begin{equation}
\label{eq:N_hat1}
     \left\langle 2^{\Delta(\tau)}\right\rangle= \left\langle \frac{v_0}{v(\tau)}\right\rangle e^\tau.
\end{equation}
This is where we can use the periodicity of the distribution of $v(\tau)$, which implies that $\langle v_0/v(\tau)\rangle$ is periodic in $\tau$ with period $\ln(2)$. We can therefore expand this function as a Fourier series
\begin{equation}\label{eq:v_Fourier}
    \left\langle \frac{v_0}{v(\tau)}\right\rangle = \sum_{k} c_k e^{ik \omega \tau},
\end{equation}
with $\omega=2\pi/\ln(2)$, where $c_k$s are constants that depend on the lineage's initial conditions and model of division. By combining Eqs.~\eqref{eq:N_hat0}, \eqref{eq:N_hat1} and \eqref{eq:v_Fourier}, we obtain
\begin{align}
    N(t) =& \sum_{k\in \mathbb{Z}} c_k \left\langle e^{\left(1+ik\omega\right) \tau_t}\right\rangle\nonumber\\
    \label{eq:N_gen_osc}
    =& e^{\langle \tau_t\rangle + \frac{1}{2}\sigma_{\tau_t}^2} \sum_{k\in\mathbb{Z}} c_k e^{ik\omega\left(\langle \tau_t\rangle +\sigma_{\tau_t}^2\right)} e^{-\frac{1}{2}k^2\omega^2 \sigma_{\tau_t}^2},
\end{align}
where $\langle \tau_t\rangle$ and $\sigma_{\tau_t}^2$ are given by \eq{tau_mean}. At large $t$ all components become negligible compared to the fastest growing term, which is $k=0$. Hence asymptotically, the population grows as
\begin{equation}
\label{eq:N_asymp0}
    N(t) \approx c_0\, e^{\langle \tau_t\rangle+\frac{1}{2} \sigma_\tau^2}.
\end{equation}
This method can also be used to find other quantities averaged over a population. If $x$ is some cell property, its value averaged over all cells in a population snapshot is 
\begin{equation}
\label{eq:pop_avg}
    \langle x\rangle_{p}=\frac{1}{N(t)}\left\langle x \frac{v_0}{v(\tau_t)}e^{\tau_t}\right\rangle,
\end{equation}
where the subscript $_p$ indicates population average (See Appendix \ref{sec:Property} for derivation).
Here $x$ can be any cell variable, such as instantaneous growth rate $\lambda_t$ or size $v(\tau_t)$. By applying \eq{pop_avg} to Dirac-delta functions $\delta(\lambda-\lambda_t)$ and $\delta(v-v(\tau_t))$ we can calculate expected population probability density distributions of growth rates and sizes respectively.\\

\section{Steady-State population properties}
Using Eqs.~\eqref{eq:tau_mean} and \eqref{eq:N_asymp0}, we find a surprisingly simple relationship for the asymptotic population growth rate
\begin{equation}
\label{eq:L_inf}
    \Lambda_\infty \equiv \lim_{t\to \infty} \frac{1}{N(t)}\frac{dN(t)}{dt}= \bar{\lambda}+\frac{\sigma_{\lambda}^2}{\theta}.
\end{equation}
When the lineage time-average cell growth rate $\bar{\lambda}$ is fixed, increasing growth rate variability increases the population growth rate. The difference between population growth rate and single-cell growth rate, given by $\sigma_\lambda^2/\theta$, can be thought of as the diffusion coefficient of accumulated lineage cell growth, since for large $t$ we have $\sigma_{\tau_t}^2 \approx 2(\sigma_\lambda^2/\theta)t$. The result that growth rate variability is always beneficial from the perspective of population growth rate seems to contradict the results of previous work~\cite{lin2017effects,lin2020single}. This difference can be explained by how growth rate variability is defined differently in the previous work. We discuss this in more detail in Section \ref{sec:Comparison}.

The distribution of instantaneous single-cell growth rates in a population snapshot can be obtained by calculating $\langle \delta(\lambda-\lambda_t)\rangle_p$ (See Appendix \ref{sec:Growth Rate Distribution} for a derivation). Asymptotically, this distribution converges to a Gaussian distribution with mean $\Lambda_\infty$ and variance $\sigma_{\lambda}^2$. Measurements of instantaneous growth rates of cells sampled from populations will thus be higher than those of cells sampled from single lineage experiments. Given that this result is again independent of the details of cell division, 
it is consistent with \rf{levien2021non} where a similar model of growth is studied neglecting noise and correlations in the division process.

\begin{figure}[t]
    \centering
    \includegraphics[scale=0.55]{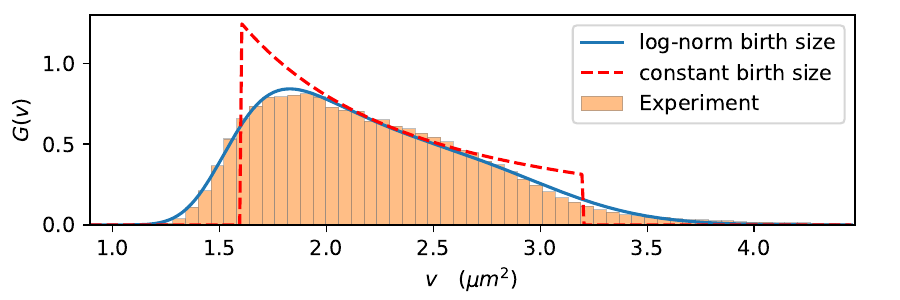}
    \caption{Comparison of the size distribution from \eq{Gx} (blue solid curve, with birth-size distribution derived from the model of cell-size regulation in \eq{csr}) to histograms of cell sizes from Ref.~\cite{gray2019nucleoid} and the inverse square law (dashed red curve, assuming deterministic division).}
    \label{fig:fig2}
\end{figure}

Next, we can calculate the time-dependent size distribution $G(t,v)=\langle \delta(v-v(\tau_t))\rangle_p$ using \eq{pop_avg}. By taking the asymptotic time limit, we find the stationary distribution of cell sizes in a population snapshot
\begin{equation}
\label{eq:Gx}
    G(v) = \frac{2}{v^2\left\langle 1/v_b\right\rangle}\left( F_{v_b}(2v)-F_{v_b}(v)\right),
\end{equation}
where $F_{v_b}$ is the cumulative distribution function of the birth size $v_b$ (see Appendix~\ref{sec:Size Distribution} for derivation). When the birth size variability approaches zero, $F_{v_b}$ becomes a step function and \eq{Gx} turns into the well-known \lq\lq inverse square law" of size distribution for constant birth size~\cite{koch1962model}, characterized by $G(v)= 2v_b/v^2$ for $v_b\leq v<2v_b$ as shown in \fig{fig2}. Remarkably, \eq{Gx} is independent of the growth process (growth rate, its fluctuations, and its correlation time), and as such, it agrees with the size distribution predicted by Powell in 1964~\cite{powell1964note}, where it was assumed that all cells grow deterministically with the same constant growth rate.

Numerous studies have examined steady-state population cell-size distributions for various models of cell growth and division. When the growth rate is modeled as a random variable that varies per cell rather than as a continuous process in time, the cell size distribution no longer exhibits decoupling from growth rate statistics~\cite{jia2021cell,nieto2021continuous}. Consequently, these models as well as alternative approaches to cell division modelling do not result in simple closed forms for population cell-size distributions~\cite{luo2021master,thomas2018analysis}. 
\\

The model of cell division described in \eq{csr} predicts a log-normal distribution $f_{v_b}$ for the birth sizes (See Appendix~\ref{sec:Birth Size} for derivation). \Fig{fig2} shows reasonable agreement between the best fit of \eq{Gx} with a log-normal $f_{v_b}$ to experimental cell sizes from \rf{gray2019nucleoid}. The slight disagreement in the tail of the size distribution hints at a potential non-Gaussianity of the noise term $\eta$ in cell division.\\

\section{The decoupling holds in the general model}
\label{sec:Decoupling}
In the last section, we saw that the asymptotic population growth rate is independent of the birth size distribution if division is symmetric and the growth rate is an Ornstein-Uhlenbeck process. In this section, we will provide a clear argument as to why this decoupling holds for any combination of the growth rate process and cell division model, including asymmetric division. We will do this by showing that the asymptotic population growth rate can be determined only from the single-cell growth process through the general relationship
\begin{equation}
\label{eq:Linf_general}
    \Lambda_\infty = \lim_{t \to\infty} \frac{1}{t} \ln\left\langle e^{\int_0^t\lambda_s ds}\right\rangle.
\end{equation}

The key argument is to consider the evolution of the expected total volume of the population, given by
\begin{equation}
\label{eq:V}
V(t) = \left\langle \sum_{j=1}^{N(t)} v_j(t)\right\rangle.
\end{equation}
Naturally, the average cell volume in the population $\langle v\rangle_p$, or equivalently the ratio of total population volume to cell count $V(t)/N(t)$ should go to a constant as the population approaches the steady state. Consequently, $V(t)$ and $N(t)$ have the same asymptotic growth rate. Interestingly, the expected total cell volume $V(t)$ is completely independent of cell divisions. Let us illustrate why we have this independence. Consider a single cell starting from volume $v(0)$ with growth rate $\lambda_0$. If there are no divisions before time $t$, then we may simply solve \eqref{eq:v_t0} to find that the total expected volume is given by
\begin{equation}
\label{eq:v0}
 V(t) = v(0) \left\langle e^{\tau_t}\right\rangle.
\end{equation}
Now suppose this cell split into two cells labeled $1$ and $2$ at $t=0$, such that the growth rate is inherited and the total cell volume is conserved $v_2(0)+v_1(0) = v(0)$. We denote the accumulated growth of the two daughter cells by $\tau^{(1)}_t$ and $\tau^{(2)}_t$ respectively. Let us now consider the expected total volume of the two daughter cells
\begin{align}
\label{eq:v1v2}
& V_1(t) + V_2(t)
= v_1(0) \left\langle e^{\tau_t^{(1)}}\right\rangle +v_2(0) \left\langle e^{\tau_t^{(2)}}\right\rangle.
\end{align}
Since any growth rate trajectories starting from the same initial growth rate $\lambda_0$ have the same distributions, the expectation values on the right-hand sides of \eq{v1v2} and \eq{v0} must all be equal. Since cell volume was conserved, we find an equality between the right-hand sides of \eq{v1v2} and \eq{v0}, and thus
\begin{equation}
    V_1(t) + V_2(t) =  V(t).
\end{equation}
We have thus shown that the expected total volume of a population stemming from one mother cell is the same as the sum of the populations stemming from its daughter cells. Using inductive arguments, we see that the total volume of any cell population as defined in \eq{V} is independent of any cell divisions. The expected value of the volume of a population stemming from one single cell with volume $v(0)$ can thus always be written as
\begin{equation}
\label{eq:EV1}
 V(t) = v(0) \left\langle e^{\tau_t}\right\rangle.
\end{equation}
Consequently, the time evolution of the expected total volume purely depends on the statistics of the growth rate process along a lineage. \eq{EV1} holds for any type of cell division mechanism, including asymmetric division. Since $V(t)$ and $N(t)$ share the same asymptotic growth rate, this rate in general must be completely independent of cell division mechanics. By matching \eq{EV1} to the definition of $\Lambda_\infty$ in \eq{L_inf}, we obtain the general expression \eq{Linf_general}. Note that the specific expression for $\Lambda_\infty$ in \eq{L_inf} could alternatively be derived using \eq{Linf_general}.

Similarly, we had shown in the previous section that the cell size distribution is independent of the statistics (mean, variance, and correlation time) of the growth process $\lambda_t$ where $\lambda_t$ is an Ornstein-Uhlenbeck process. In Appendix \ref{sec:Size Distribution}, we show that this decoupling holds for more general models of growth and division. The cell size distribution given in \eq{Gx} holds for arbitrary growth process $\lambda_t$ and birth size distributions $F_{v_b}(v)$.

\section{Discrete generational growth from course-graining continuous growth}
\label{sec:Comparison}
In both experimental and theoretical studies, the single-cell growth rate is often defined as the generational growth rate of a cell averaged over its lifetime~\cite{stewart2005aging,robert2014division,taheri2015cell,lin2017effects,lin2020single,campos2014constant,jia2021cell}. In this section, 
we will coarse-grain the growth fluctuations over the cell cycle to derive an effective model of growth consistent with existing literature. 
While this coarse-graining keeps the qualitative behavior the same, it introduces subtle dependencies between otherwise decoupled processes of growth and division as observed in \rf{lin2020single,jia2021cell,barber2021modeling}. Let us denote the coarse-grained generational growth rate by the variable $\kappa$
\begin{equation}
    \label{eq:kappa_def}
        \kappa \equiv \frac{1}{t_d-t_b} \int_{t_b}^{t_d} \lambda_s ds,
\end{equation}
where $t_b$ and $t_d$ are birth and division times, the latter being defined as a passage time of the cell reaching the target division size $v_d$, (see Appendix \ref{sec:Comparison2} for a more precise definition).

In the small noise limit where $\sigma_\lambda \ll \bar\lambda$, we can link the mean $\bar\kappa$ and variance $\sigma_\kappa^2$ of the effective generational growth rate to the underlying growth process's parameters $\bar\lambda$, $\sigma_\lambda$ and $\theta$. In Appendix \ref{sec:Comparison2} we show that up to first order in $\sigma_\lambda^2/\bar\lambda^2$, we have
\begin{equation}
\label{eq:mkappa}
    \bar{\kappa} \approx \bar\lambda+ \frac{\sigma_\lambda^2}{\bar\lambda} h( \gamma), \qquad \sigma_{\kappa}^2\approx \sigma_\lambda^2 h(\gamma).
\end{equation}
Here $h(\gamma)$ is some dimensionless function of the ratio of correlation decay rate to growth rate set by $\gamma = \ln(2)\theta/\bar\lambda = t_{div}/t_{corr}$. $h(\gamma)$ is given by
\begin{equation}
\label{eq:hgam}
    h(\gamma) = \frac{2}{\gamma}\left(1-\frac{1}{\gamma}\left(1-e^{-\gamma}\right)\right).
\end{equation}
When correlations are strong $\gamma \ll 1$, growth rate $\lambda_t$ is approximately constant over the lifetime of a cell and we find that $h(\gamma) \approx 1$. When correlations vanish $\gamma \gg 1$, we obtain $h(\gamma) \approx 2/\gamma$. 
We can now start rewriting equation \eq{L_inf} in terms of effective generational growth rate parameters
\begin{equation}
\label{eq:L_kappa}
    \Lambda_\infty \approx \bar\kappa + \left(\frac{\ln(2)}{\gamma h(\gamma)} - 1\right) \frac{\sigma_\kappa^2}{\bar\kappa}.
\end{equation}
For vanishing mother-daughter growth rate correlations, \eq{L_kappa} recovers the result from references~\cite{lin2017effects,lin2020single,jafarpour2019cell}. An interesting observation is that $\Lambda_\infty$ can be either larger or smaller than $\bar\kappa$, depending on the strength of the correlations. The crossover where $\Lambda_\infty =\bar\kappa$ occurs at intermediate values when $\theta\approx 0.64 \bar\lambda$. This is in close agreement with the result from \rf{lin2020single}, where a similar crossover was found that depends on the mother-daughter generational growth rate correlations. Based on \rf{lin2020single} or \eq{L_kappa} alone, one may be tempted to think growth rate variability is either beneficial or disadvantageous to population growth, depending on the correlation strength. In the continuous model of growth, when $\bar \lambda$ is kept constant, the growth fluctuations are always beneficial to the population growth rate $\Lambda_\infty$. The cross-over observed in the coarse-grained model arises upon fixing the coarse-grained variable $\bar \kappa$ which itself depends on the growth correlations.\\

It is worth noting that in the presence of non-vanishing growth rate correlations, the models of continuous growth and generational growth are not completely equivalent. This is best illustrated by looking at the correlations of cells across generations. For a cell $i$ with effective growth rate $\kappa_i$ and a direct offspring $n$ generations removed with effective growth rate $\kappa_{i+n}$, we define the generational correlation function as $\rho_n = \text{Cov}(\kappa_{i},\kappa_{i+n})/\sigma_\kappa^2$. In a coarse-grained model where $\kappa_i$s are defined as a Markov chain, the correlation function would have to satisfy $\rho_n = (\rho_1)^n$. When the effective growth rates are based on an underlying continuous process by \eq{kappa_def}, the generational correlations $\rho_n$ will follow a different pattern, discussed in Appendix \ref{sec:Comparison2}. In \fig{rho} we show an example of $\rho_n$ when fixing $\rho_1$ and $\sigma_\kappa \ll \bar\kappa$

\begin{figure}[t]
    \centering
    \includegraphics[scale=0.70]{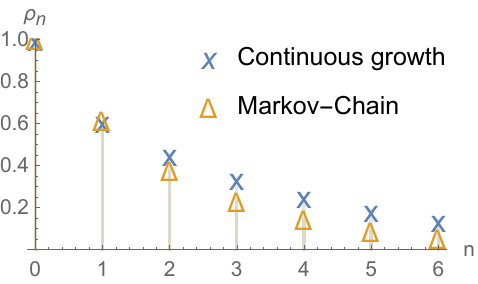}
    \caption{A comparison of the generational growth rate correlation $\rho_n = Cov(\kappa_i,\kappa_{i+n})$ between cells that are $n$ generations removed.We fix $\rho_1$ and assume $\sigma_\kappa \ll \bar\kappa$. In blue, we show $\rho_n$ given by \eq{rho_n}, which are the generational correlations if $\kappa_i$ emerge from some underlying continuous growth rate. In yellow, we show $\rho_n=(\rho_1)^n$, which are the growth rate correlations when $\kappa_i$ are modeled as a simple Markov Chain.}
    \label{fig:rho}
\end{figure}

In the presence of division noise, there are small corrections to Eqs.~\eqref{eq:mkappa} and \eqref{eq:L_kappa} of order $\sigma_\kappa^2\sigma_{v_b}^2$ (also observed in Refs.~\cite{lin2020single,thomas2017single}) which prevent the decoupling between population growth rate and the division process when the model is expressed in terms of these variables.

\section{Transient Dynamics of population properties}
The dynamics of the number of cells in the population is extremely well approximated by only the zeroth and first-order terms from \eq{N_gen_osc} and the large time behavior of $\sigma_{\tau_t}^2$ and $\langle \tau_t\rangle$ given by \eq{tau_mean}. This results in population dynamics of the form
\begin{align}
\label{eq:N_osc1}
   N(t) \approx C e^{\Lambda_\infty t}\left(1 + A  Cos\left(\Omega t+ \phi  \right) e^{-r  t}\right),
\end{align}
where $\Omega =\omega\left(\bar{\lambda} + 2\sigma_{\lambda}^2/\theta\right)$, $r = \omega^2\sigma_{\lambda}^2/\theta$ with $\omega = 2\pi/\ln(2)$ (the constants $C$, $A$, and $\phi$ can be calculated explicitly as functions of the initial state of the population; see Appendix \ref{sec:Size Distribution} for the exact solution). \Eq{N_osc1} tells us that the population growth rate (defined as $\Lambda(t)\equiv d\ln(N)/dt$) oscillates, and its amplitude decays exponentially with rate $r$. The uncorrelated limit ($\theta\to\infty$ keeping $\sigma^2/\theta$ constant) of these oscillations is studied in \rf{jafarpour2019cell}. The numerically inferred relationship for the decay rate of these oscillations 
in \rf{jafarpour2019cell} is in perfect agreement with our analytical prediction for $r$.

\begin{figure}[t]
    \centering
    \includegraphics[scale=1.2]{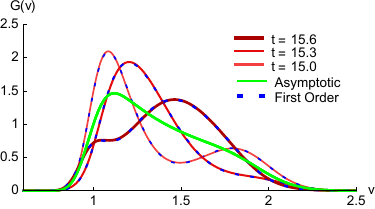}
    \caption{Traveling waves in the dynamics of size distributions: analytically calculated population size distributions are shown at three consecutive times (red solid curves) compared with the first-order approximations (blue dashed curves) and the asymptotic size distribution (green solid curve). Times are given in units of average division times $\ln(2)/\bar{\lambda}$. Model parameters: $\theta =\bar{\lambda}=\ln(2)$, $\sigma_\lambda^2/\theta= 0.006 \bar{\lambda}$, $\sigma_{\ln(v_b)}^2=0.2$. The time evolution of size distribution is shown in the Supplemental Material~\cite{SM}.}
    \label{fig:Gplot}
\end{figure}

The time-dependent population cell size distribution can be calculated explicitly by taking $\langle \delta(v-v(\tau_t))\rangle_p$. The first-order large-time approximation becomes 
\begin{align}
\label{eq:Gv1}
    G(t,v) \approx  G(v)\frac{1+BCos\left(\Omega t -\omega \ln(v) + \zeta \right)e^{-r  t}}{1+A  Cos(\Omega  t + \phi )e^{-r  t}},
\end{align}
where $A$, $B$, $\phi$ and $\zeta$ depend on the initial conditions of the population, and $G(v)$ is the asymptotic size distribution given by \eq{Gx} (see Appendix \ref{sec:Size Distribution} for the exact solution).
A plot of \eq{Gv1} is given in \fig{Gplot}, where we compare it to the exact time-dependent size distribution and the asymptotic case.
We see that the deviation from the asymptotic distribution decays at the rate $r$.
The parameter $r $ therefore provides us with a timescale for the internal population properties to reach the steady state. In \eq{Gv1} we also encounter traveling waves in logarithmic cell size densities. These waves propagate at a speed of $\Omega /\omega=\bar{\lambda}+2\sigma_\lambda^2/\theta$. Interestingly, this is faster than the population growth rate $\Lambda_\infty$ from \eq{L_inf}.\\

\section{Discussion}
Previous works studying the role of growth variability in population dynamics consider the growth rate as a constant random variable associated with each cell and ignore its continuous fluctuations over time~\cite{lin2017effects,lin2020single,jafarpour2019cell,taheri2015cell,ho2018modeling,marantan2016stochastic,barber2021modeling}. Here, we have shown that if instead of simplified models, we consider a less coarse-grained more accurate model of single-cell growth, the roles of growth and division statistics on population dynamics decouple, and all aspects of population dynamics become analytically solvable.

For the cell size distribution, we have recovered a classical result in \eq{Gx} and after nearly 60 years since its discovery~\cite{powell1964note}, we have shown that it also holds for the case with fluctuating growth rates. This result implies that a snapshot image of a population of cells cannot be used to extract any information about either the growth process or the mechanism of cell size regulation. However, our time-dependent solution \eq{Gv1} predicts traveling waves in population size distribution that can be used to infer single-cell growth rate variability.

For the population growth rate, we discovered a remarkably simple relationship in \eq{L_inf}. This equation connects the statistics of growth fluctuations (and therefore fluctuations in concentrations of the underlying molecular components) to the fitness of the organism. This allows for studying (independently of any model of cell division) how these fluctuations at the molecular level are shaped by evolution~\cite{jafarpour2023evolutionary}.

Additionally, Eqs.~\eqref{eq:N_osc1} and~\eqref{eq:Gv1} for the first time, provide closed-form solutions for the transient dynamics of the population with realistic models of growth and division.

\begin{acknowledgments}
We gratefully acknowledge Philipp Thomas and Ethan Levien for their comments on the manuscript and Anna Ochab-Marcinek, Marcin Rubin, Jakub J\k{e}drak, and Piotr Wi\k{a}cek for insightful discussions. This work is part of the D-ITP consortium, a program of the Netherlands Organisation for Scientific Research (NWO) that is funded by the Dutch Ministry of Education, Culture and Science (OCW).
\end{acknowledgments}

\appendix

\section{Birth size}
\label{sec:Birth Size}
We defined the division size $v_d$ of a cell to depend on its birth size $v_d$ via \eq{csr}. We denote a $n$-th generation cell's birth and division size by $v_{b,n}$ and $v_{d,n}$ respectively. Given that cells divide symmetrically, a cell's birth size would be half the size of its mother cell at division, hence we have that $v_{b,n}=v_{d,n-1}/2$. We can now rewrite \eq{csr} as a birth size recursion formula
\begin{equation}
    \ln(v_{b,n+1}/\bar v) = (1-\alpha) \ln(v_{b,n}/\bar v) + \eta_n.
\end{equation}
We can solve this to express any cell's birth size in terms of the first cell's birth size $v_{b,0}$ and all noise terms
\begin{equation}
    \ln(v_{b,n}/\bar v) = (1-\alpha)^n\ln(v_{b,0}/\bar v) + \sum_{i=0}^{n-1}\eta_i (1-\alpha)^{n-1-i}.
\end{equation}
As $n$ increases, the birth size distribution rapidly converges to some steady state distribution, with mean and variance
\begin{equation}
\label{eq:vbn_meanvar}
    \langle \ln v_{b,n}\rangle = \ln \bar v, \qquad \sigma^2_{\ln v_{b,n}} = \frac{1}{\alpha(2-\alpha)} \sigma_\eta^2.
\end{equation}
For all but very early generations, it is safe to assume that $\ln v_{b,n}$ is normally distributed with mean and variance given by \eq{vbn_meanvar}.

\section{Ornstein-Uhlenbeck process}
\label{sec:OU process}
In the main text, the growth rate was defined by \eq{l_ou}. Its general solution can be expressed as
\begin{equation}
       \lambda_t = \bar{\lambda} + (\lambda_0-\bar{\lambda})e^{-\theta t} + \sigma_\lambda \sqrt{2\theta} e^{-\theta t}\int_0^t e^{\theta s} \xi(s)ds.
\end{equation}
Due to the properties of $\xi(t)$, upon fixing $\lambda_0$, growth rate $\lambda_t$ is Gaussian with mean
\begin{align}
\label{eq:lambda_mean}
        \langle \lambda_t \rangle = \bar\lambda + \left(\lambda_0 -\bar\lambda \right) e^{-\theta t}
\end{align}
and covariance
\begin{equation}
    \label{eq:lambda_covariance}
    \text{Cov}(\lambda_t,\lambda_s) = \sigma_\lambda^2 \left(e^{-|t-s|}-e^{-\theta| t+s|}\right).
\end{equation}
As a consequence, the accumulated growth defined in 
\eq{tau_def0} is also Gaussian. It can be calculated by plugging in \eq{lambda_mean} and \eq{lambda_covariance}. Up to initial condition-dependent constants we have that the mean is given by
\begin{equation}
    \langle \tau_t \rangle = \int_0^t \langle\lambda_s\rangle ds = \bar \lambda t + \frac{1-e^{-\theta t}}{\theta}(\lambda_0-\bar\lambda),
\end{equation}
and the variance is given by
\begin{align}
    \sigma_{\tau_t}^2 =& \int_0^t \int_0^t \text{Cov}(\lambda_s,\lambda_u)dsdu \\\nonumber =&\frac{\sigma_\lambda^2}{\theta} t - \frac{\sigma_\lambda^2}{\theta^2} \left(3 - 2e^{-\theta t} + e^{-2\theta t}\right).
\end{align}
Upon removing the exponential corrections, we obtain the long-term linear behavior as in \eq{tau_mean}

\section{Population property densities}
\label{sec:Property}
Let $x^j_t$ be some value related to cell $j$, like cell size $v^j(t)$ or growth rate $\lambda^j_t$. We define the total value $\tilde X(t)$ as the sum of $x^j_t$ over all cells in the population at time $t$. This sum can be obtained by summing over $x_t^j$ over all cells within the population tree but with $x^j_t$ multiplied by zero if $j$ is not in the population and by one otherwise. We can thus write
\begin{equation}
    \tilde X(t) = \sum_{n=0}^\infty \sum_{j=1}^{2^n} x_t^j \delta_{\Delta^j(\tau^j_t),n}.
\end{equation}%
Since all lineages are equal in distribution, we find that the expected total sum of $x^j_t$ over all cells is given by
\begin{equation}
    X(t) = \left\langle \tilde{X}(t)\right\rangle = \sum_{n=0}^\infty 2^n \left\langle x_t \delta_{\Delta(\tau_t),n}\right\rangle,
\end{equation}%
where in the last equality we again used the fact that the distributions of cell properties are lineage-independent. For each term, we can now put the factor two within the expectation brackets and replace $n$ by $\Delta(\tau_t)$ due to the presence of the Kronecker delta. We obtain
\begin{equation}
\label{eq:Xt}
    X(t) = \sum_{n=0}^\infty \left\langle x_t 2^{\Delta(\tau_t)}\delta_{\Delta(\tau_t),n}\right\rangle = \left\langle x_t 2^{\Delta(\tau_t)}\right\rangle.
\end{equation}%
The average value of $x_t$ per cell in the population can now simply be obtained by dividing this total by the expected number of cells, hence
\begin{equation}
\label{eq:xp}
    \langle x_t\rangle_p = \frac{X(t)}{N(t)} = \frac{1}{N(t)} \left\langle x_t \frac{v_0}{v(\tau_t)} e^{\tau_t}\right\rangle.
\end{equation}%
By setting $x_t$ to $\delta(\lambda_t-\lambda)$ or $\delta(v(\tau_t)-v)$ one can calculate the time-dependent distributions of $\lambda_t$ and $v(\tau_t)$ respectively. 

\section{The periodicity of lineage cell size}
\label{sec:Periodicity}
In \ref{sec:Derivation} it was argued that the distribution of cell size $v(\tau)$ was periodic in $\tau$ without proof. In this section, we provide such a proof. The idea is that when $\tau$ increases by $\ln(2)$, the cell size $v(\tau)$ will double in size but also halve either $0$, $1$, or $2$ times depending on the division sizes. When cell division sizes are in steady-state, we shall see that these odds of division must be balanced such that the distribution of $v(\tau)$ remains unaltered upon increasing $\tau$ by $\ln(2)$. Recall \eq{vrel} for cell size as a function of stochastic time 
\begin{equation}
\label{eq:vrel2}
    v(\tau) = \frac{v_0 e^{\tau}}{2^{\Delta(\tau)}}.
\end{equation}

Assuming that $n$ divisions have already occurred, the next division occurs when the cell's size $v(\tau)=v_0e^{\tau}/2^n$ crosses the cell's division size, given by $2v_{b,n+1}$. The total number of divisions $\Delta(\tau)$, given some stochastic time $\tau$ can thus be obtained by checking for how many generations $n$, this condition is satisfied, hence
\begin{equation}
\label{eq:Delta_def}
    \Delta(\tau) = \#\{n\geq 0: v_0e^\tau/2^n \geq 2v_{b,n+1}\}.
\end{equation}
We assume that for large $n$, the birth sizes $v_{b,n}$ are identically distributed with some steady-state distribution given by $v_b$. After dropping this index, we can rewrite \eq{Delta_def} to an expression identical in distribution
\begin{equation}
\label{eq:Delta_sim}
    \Delta(\tau) \sim \left\lfloor \frac{\tau+\ln(v_0/v_b)}{\ln(2)}\right\rfloor,
\end{equation}
where $\lfloor.\rfloor$ denotes the floor function. Consequently, using \eq{vrel2} we find that the log cell size distribution follows
\begin{equation}
   \ln\left( v(\tau)\right) \sim \ln(v_0)+ \tau - \left\lfloor \frac{\tau+\ln(v_0/v_b)}{\ln(2)}\right\rfloor \ln(2).
\end{equation}
Note how the right-hand side remains unaltered upon a shift of $\tau \to \tau+\ln(2)$, hence the distribution of $v(\tau)$ is periodic in $\tau$ with a period of $\ln(2)$.

\section{Size distribution}
\label{sec:Size Distribution}
The size distribution can be calculated as 
\begin{equation}
\label{eq:G_def}
    G(t,v) = \frac{\left\langle \delta(v(t) -v) 2^{\Delta_t} \right\rangle}{N(t)}.
\end{equation}
Conditioned to $v(t)=v$, the number of divisions $\Delta(\tau_t)$ can be rewritten in terms of current and accumulated size as $2^{\Delta(\tau_t)} = v_0 e^{\tau_t}/v$, hence
\begin{equation}
\label{eq:G_def1}
G(t,v) = \frac{v_0}{v}\frac{1}{N(t)} \left\langle \delta\left(v(t)-v\right) e^{\tau_t}\right\rangle.
\end{equation}

Furthermore, using \eq{vrel} and \eq{Delta_sim} we can rewrite
\begin{align}
     \delta(v(t) - v) \sim \frac{1}{v} \delta\left(  \tau_t^* - \left\lfloor \frac{\tau_t^*}{\ln 2} \right\rfloor\ln(2) - \ln(v/v_b)\right),
\end{align}
where
\begin{equation}
    \tau_t^* = \tau_t + \ln(v_0/v_b).
\end{equation}
Note $\tau_t^*-\ln(2)\lfloor \tau_t^*/\ln(2) \rfloor$ is a sawtooth function in $\tau_t^*$ that ranges from $0$ to $\ln(2)$. The delta function thus evaluates to zero when $\ln(v/v_b)$ does not lie within $[0,\ln(2)]$. For constant birth size $v_b$, this is easy to interpret as the condition that $v$ is either larger than the maximum size $2v_b$ or smaller than the minimum size of $v_b$. For these values of $v$ the population cell size distribution should indeed have zero density. If $v$ does, however, fall within the allowed range, we get that $\delta(v(t)-v)$ is periodic in $\tau_t$, with peaks at $\tau_t=\ln(v/v_0)$, hence we may Fourier expand this as
\begin{equation}
    \delta(v(t)-v) \sim \frac{\mathbf 1\{v_b\leq v\leq 2v_b\}}{v\ln(2)} \sum_{k=-\infty}^\infty e^{i \omega k (\tau_t-\ln(v/v_0))},
\end{equation}
where $\omega = 2\pi/\ln(2)$. As we plug this expression back into \eq{G_def1}, we see that the dependency on birth size distribution factors out, and we are left with
\begin{align}
\label{eq:Gv_full}
    G(t,v) =&  \frac{2 v_0}{v^2} \mathbb P \left( v_b\leq v<2v_b\right)  \notag\\
    & \times\frac{\sum_{k=-\infty}^\infty e^{i\omega k \ln(v_0/v)} \left\langle e^{(1+i\omega k ) \tau_t}\right\rangle}{\sum_{k=-\infty}^\infty c_k \left\langle e^{(1+i\omega k ) \tau_t}\right\rangle}.
\end{align}

In \eq{Gv_full} both the numerator and denominator, all transient terms, which are $\left\langle e^{(1+i\omega k)\tau_t}\right\rangle$ for $k\neq 0$, vanish as $t$ goes to infinity. The asymptotic part is thus given by \eq{Gx}. If we truncate the sum numerator and denominator in \eq{Gv_full} at the same value of $|k|$ we obtain a properly normalized size distribution function that becomes an increasingly better approximation of the exact time-dependent size distribution as $t$ becomes large. Truncating at $|k|\leq 1$ yields \eq{Gv1}. 
\\

To show that the full time-dependent solution in \eq{Gv_full} converges the steady-state size distribution in \eq{Gx}, we look at the asymptotic value of the ratio of the higher order terms $k \neq 0$ relative to the leading term $k=0$,
\begin{equation}
\label{eq:cond}
   \lim_{t\to\infty} \frac{\langle e^{(1+i\omega k)\tau_t}\rangle}{\langle e^{\tau_t}\rangle}.
\end{equation}
The numerator of this expression is a rapidly oscillating integral as $t$ approaches infinity. Hence, for any reasonable growth process $\lambda_t$, we expect this ratio to approach zero, as we could explicitly confirm for the case of an Ornstein-Uhlenbeck process. When this condition is satisfied, all the higher-order terms drop out and the steady-state size distribution converges to \eq{Gx}.

\section{The transient amplitude and phase}
\label{sec:Amplitude}
In this section, we discuss how to obtain expressions for the Fourier constants $c_k$ in the expansion in \eq{v_Fourier} and use those to obtain expressions for constants in transient population count and size distribution in \eq{N_osc1} and \eq{Gv1} respectively. From the very definition of size distribution in \eq{G_def} we know that $G(t,v)$ must be normalized over $v$. Consequently \eq{Gv_full} must be a normalized probability density function over $v$. We can use this to obtain the Fourier coefficients $c_k$ of $N(t)$ as they appear in \eq{N_gen_osc} as
\begin{equation}
    c_k = \frac{1}{1+i\omega k} v_0^{1+i\omega k} \left\langle 1/v_b^{1+i\omega k} \right\rangle.
\end{equation}
If the steady-state birth size distribution $v_b$ is log-normal with parameters given by \eq{vbn_meanvar}, then these coefficients can be evaluated as
\begin{equation}
    c_k = \frac{1}{1+i\omega k} (v_0/\bar v)^{1+i\omega k} e^{\frac{1}{2}\left( 1-\omega^2 k^2 +i2\omega k\right) \sigma_{\ln v_b}^2}.
\end{equation}
The first order behavior in \eq{N_osc1} written as
\begin{equation}
    N(t) \approx C e^{\Lambda_\infty t}\left(1+ A\cos(\Omega t+ \phi)e^{-qt}\right).
\end{equation}
We want to match this to the behavior of \eq{N_gen_osc} truncated at $|k|=1$
\begin{align}
    N(t) &\approx  e^{\langle \tau_t \rangle + \frac{1}{2} \sigma_{\tau_t}^2}\\ \nonumber  \times &\left(c_0+ 2\text{Re}\left[c_1 e^{ik\omega \left(\langle\tau_t\rangle+\sigma_{\tau_t}^2\right)}e^{-\frac{1}{2} k^2 \omega^2 \sigma_{\tau_t}^2}\right]\right).
\end{align}
If we incorporate the growth lag constants in stochastic time given in \eq{tau_mean}, we find a population constant 
\begin{equation}
    C =   \frac{v_0}{\bar v}  e^{ \frac{1}{2} \sigma_{\ln v_b}^2+\frac{\lambda_0-\bar\lambda}{\theta} - \frac{3}{2}\frac{\sigma_\lambda^2}{\theta^2}},
\end{equation}
an oscillation amplitude
\begin{equation}
     A = \frac{2}{\sqrt{1+\omega^2}} e^{-\frac{3}{2} \omega^2 \frac{\sigma_\lambda^2}{\theta^2}},
\end{equation}
and a phase
\begin{align}
    \phi =& \omega\left(\ln(v_0/\bar v) + \sigma_{\ln v_b}^2 + \frac{\lambda_0-\bar\lambda}{\theta} - 3\frac{\sigma_\lambda^2}{\theta^2}\right)\\
    & + \arctan(\omega).
\end{align}
The constants $B$ and $\zeta$ in the transient size distribution now similarly follow as
\begin{equation}
    B = 2e^{-\frac{3}{2} \omega^2 \frac{\sigma_\lambda^2}{\theta^2}}
\end{equation}
and
\begin{equation}
    \zeta = \omega\left( \ln(v_0)+ \frac{\lambda_0-\bar\lambda}{\theta} - 3\frac{\sigma_\lambda^2}{\theta^2}\right).
\end{equation}

\section{Growth rate distribution}
\label{sec:Growth Rate Distribution}
The growth rate distribution can be calculated as 
\begin{equation}
\label{eq:rho_def}
    \rho(t,\lambda) = \frac{\left\langle \delta(\lambda_t - \lambda) 2^{\Delta_t}\right\rangle}{N(t)}.
\end{equation}
By performing a set of operations similar to the ones used to obtain \eq{Gv_full}, we may rewrite \eq{rho_def} as
\begin{equation}
    \rho(t,\lambda) = \frac{\sum_{k=-\infty}^\infty \left\langle \delta(\lambda_t-\lambda)e^{(1+i\omega k)\tau_t^*}\right\rangle}{\sum_{k=-\infty}^\infty \left\langle e^{(1+i\omega k)\tau_t^*}\right\rangle}
\end{equation}
to find the asymptotic growth rate distribution, we let $t$ go to infinity. In this case, one finds that all the transient terms disappear, and we are left with
\begin{equation}
    \rho(\lambda) = \lim_{t\to\infty}\frac{\left\langle \delta(\lambda_t-\lambda)e^{\tau_t}\right\rangle}{\left\langle e^{\tau_t}\right\rangle}.
\end{equation}
To further solve this, we may calculate the distribution of $\tau_t$ conditioned to $\lambda_t=\lambda$ using \eq{lambda_covariance} and the normality of $\lambda_t$ and $\tau_t$. We find that
\begin{equation}
    \rho(\lambda) = \frac{1}{\sqrt{2\pi \sigma_\lambda^2}} e^{-\frac{(\lambda-\Lambda_\infty)^2}{2\sigma_\lambda^2}}.
\end{equation}
In other words, the instantaneous growth rate in a steady-state population is normally distributed around $\Lambda_\infty$
\section{Comparison to generational growth rate models}
\label{sec:Comparison2}
In this section we are interested in the mean and variance of the cell averaged growth rate $\kappa$ as defined in \eq{kappa_def}. Let us pick a random cell along a lineage and fix the birth size $v_b$ and division size $v_d$. For simplicity, we set $t_b=0$ at the time of this cell's birth. From \eq{v_t0} it follows that the division size satisfies
\begin{equation}
    v_d = v(t_d) = v_b e^{\int_0^{t_d} \lambda_s ds}.
\end{equation}
We define the target log-size as $u:=\ln(v_d/v_b)$
The cell's lifetime $T_u=t_d-t_b$ is now set as the hitting time $T_u$ such that
\begin{equation}
\label{eq:urel}
    u = \int_0^{T_u} \lambda_s ds.
\end{equation}
The cell-averaged growth rate $\kappa$ can then be related to this division time as
\begin{equation}
\label{eq:kappa_u}
    \kappa = \frac{u}{T_u}.
\end{equation}
Before we can derive the mean and variance of $T$ and $\kappa$ we need to consider what the initial growth rate of our cell is. One may be tempted to think that the growth rate at birth $\lambda_0$ of a randomly selected cell is distributed according to the growth rate steady-state distribution of $N(\bar\lambda, \sigma_\lambda^2)$. This is not completely right. Since division is triggered by cell size in our model, growth rate measurements of newborn cells will follow the distribution of growth rate measured at a predetermined cell size.
If one samples cell growth rate at a predetermined cell size rather than at a predetermined moment in time, there will be a bias favoring faster-growing cells. To illustrate this, consider what happens when cells switch between growing with rate $\lambda_t=\bar \lambda$ and not growing at all with rate $\lambda_t=0$ half of the time. Measuring the instantaneous growth rate of a newborn cell at its birth must yield $\lambda_t=\bar\lambda$ with full probability since its mother cell must have been growing to trigger the division in the first place. Let us formalize this argument and find the growth rate distribution of a newborn cell.

Denote by $f^*(\lambda)$ the growth rate distribution of $\lambda_{T_u}$ of $u$ uniformly selected from log-cell size space, which is equivalent to the growth rate of a random newborn cell. We have that
\begin{equation}
    f^*(\lambda)d\lambda \propto  \left\langle \int du \mathbf 1\left\{ \lambda_{T_u} \in [\lambda ,\lambda+d\lambda]\right\} \right\rangle.
\end{equation}
For positive growth rates, $u=\int_0^t\lambda_sds$ is a strictly increasing function, so we can substitute $t$ for $u$, with $du = \lambda_t dt$ to find that
\begin{equation}
    f^*(\lambda) d\lambda \propto \left\langle \int dt \lambda_t \mathbf 1\left\{ \lambda_t \in [\lambda,\lambda+d\lambda]\right\}\right\rangle.
\end{equation}
On the right-hand side, $\lambda_t$ can be set to $\lambda$ and taken out of the integral. What remains in the integral is the distribution of $\lambda_t$ sampled from uniformly selected time. By further imposing that $f^*(\lambda)$ is normalized properly, we find
\begin{equation}
\label{eq:f_asterisk}
    f^*(\lambda) = \frac{\lambda}{\bar\lambda} f(\lambda).
\end{equation}
This will now be the probability density distribution of the growth rate at birth of our randomly selected cell. The cell lifetime can now be expressed as
\begin{equation}
    T_u = \int_0^u \frac{d T_w}{dw} dw =  \int_0^u \frac{1}{\lambda_{T_w}}dw.
\end{equation}
Note that $\lambda_{T_w}$ follows the size-selected growth rate distribution, hence $\left\langle 1/\lambda_{T_w}\right\rangle = 1/\bar\lambda$. Consequently, we find that average division times simply scale with the reciprocal time-average growth rate $\langle T_u\rangle = u/\bar\lambda$.

Let us now consider small deviations around this mean value denoted by $\delta T = T_u - u/\bar\lambda$. We also denote growth rate deviations from the mean by $\delta \lambda_t = \lambda_t-\bar\lambda$. We can now rewrite \eq{urel} as
\begin{equation}
\label{eq:urel2}
    0= \int_0^{u/\bar\lambda} \delta \lambda_s ds +\delta T \bar\lambda + \int_{u\bar\lambda}^{u/\bar\lambda +\delta T} \delta \lambda_s ds.
\end{equation}
We are interested in $\langle \delta T^2\rangle$ up to first order in $\sigma_\lambda^2$, so we may ignore the rightmost term in \eq{urel2} and set
\begin{equation}
\label{eq:delta_T2}
    \left\langle \delta T^2\right\rangle = \frac{1}{\bar\lambda^2} \left\langle \left(\int_0^{u/\bar\lambda} \delta \lambda_sds\right)^2\right\rangle.
\end{equation}
By applying \eq{lambda_mean} \eq{lambda_covariance} and by letting $\lambda_0$ be distributed as in \eq{f_asterisk}, we find that
\begin{equation}
\label{eq:delta_T2h}
    \langle \delta T^2\rangle = \frac{u^2\sigma_\lambda^2}{\bar\lambda^4} h(\gamma),
\end{equation}
where $\gamma= u\theta/\bar\lambda$ and $h(\gamma)$ is given by
\begin{equation}
    h(\gamma) = \frac{2}{\gamma}\left(1-\frac{1}{\gamma}\left(1-e^{-\gamma}\right)\right).
\end{equation}
By expanding \eq{kappa_u} in small $\delta T$ and taking the expectation value, we find
\begin{equation}
    \bar\kappa =\bar\lambda \left\langle \frac{1}{1+\frac{\delta T}{u/\bar\lambda}}\right\rangle \approx \bar\lambda \left(1+ \frac{\bar\lambda^2}{u^2} \langle \delta T^2\rangle\right),
\end{equation}
and
\begin{equation}
\label{eq:sigma_kappa}
    \sigma_\kappa^2 = \langle\kappa^2\rangle-\bar\kappa^2 \approx \frac{\bar\lambda^4}{u^2} \langle \delta T^2\rangle,
\end{equation}
hence we have
\begin{equation}
    \bar\kappa \approx \bar\lambda\left(1+\frac{\sigma_\lambda^2}{\bar\lambda^2}h(\gamma)\right),\quad \sigma_\kappa^2 \approx \sigma_\lambda^2 h(\gamma),
\end{equation}
which result in \eq{mkappa}, if we choose $u=\ln(2)$, corresponding to deterministic division after doubling cell size. In the presence of birth- and division-size noise, $u$ and $\gamma$ become random variables. Assuming noise in division noise is small, we can write the mean cell-averaged growth rate as
\begin{equation}
    \bar\kappa \approx \bar\lambda\left(1+ \frac{\sigma_\lambda^2}{\bar\lambda^2} \left(h(\langle\gamma\rangle) + \frac{1}{2}\sigma_\gamma^2h''(\langle\gamma\rangle)\right)\right),
\end{equation}
hence the difference between cell-averaged growth rate $\bar\kappa$ and time-averaged growth rate $\bar\lambda$ is weakly coupled to division size noise. Consequently, any attempt to express the population growth rate $\Lambda_\infty$ in terms of cell-averaged growth rate parameters $\bar\kappa$ and $\sigma_\kappa^2$ must also exhibit weak coupling to $\sigma_\gamma^2$, proportional to division size variability. This explains the weak coupling of the population growth rate to division size variability observed in \rf{lin2020single}, despite the coupling's absence in our continuous growth rate model.

\subsubsection{Mother-daughter correlations}
In this section we show how to derive the generational correlation function $\rho_n=\text{Cov}(\kappa_i,\kappa_{i+n})/\sigma_\kappa^2$ if $\kappa_i$ and $\kappa_{i+n}$ are emergent effective growth rates of cells removed $n$ generations along a single lineage. For simplicity, we will assume constant cell division size noise, so $u=\ln(2)$, but the following arguments can easily be generalized to allow for noise and correlations in division size. For a cell $i$ with lifetime $T_i$, we can write a small growth noise expansion
\begin{equation}
    \kappa_i \approx  \bar\lambda \left( 1- \frac{\bar\lambda \delta T_i}{u} + \frac{\bar\lambda^2 \delta T_i^2}{u^2}\right)
\end{equation}
hence the small noise correlation function becomes
\begin{equation}
    \rho_n = \frac{1}{\sigma_\kappa^2} \frac{\bar\lambda^4}{u^2} \left\langle \delta T_0 \delta T_{n}\right\rangle.
\end{equation}
In order to derive the cross-term between the hitting times of subsequent cells, we can use first-order approximations to derive an expression similar to \eq{delta_T2} 
\begin{align}
    &\langle \delta T_0\delta T_n\rangle \\ \nonumber
    =&\frac{1}{\bar\lambda^2} \left\langle \left(\int_0^{u/\bar\lambda}\delta \lambda_sds\right)\left(\int_{n u/\bar\lambda}^{(n+1)u/\bar\lambda} \delta \lambda_{s'}ds'\right)\right\rangle.
\end{align}
For $n\geq 1$ we find
\begin{equation}
\label{eq:rho_n}
    \rho_n \approx \frac{1}{2} \frac{\left(1-e^{-\gamma}\right)^2}{\gamma-\left(1-e^{-\gamma}\right)}e^{-(n-1)\gamma}.
\end{equation}

\bibliography{ref}

\end{document}